# Inhomogeneous Thermal Conductivity Enhances Thermoelectric Cooling

Tingyu Lu,[1] Jun Zhou,[1,]* Nianbei Li,[1] Ronggui Yang,[2] and Baowen Li[1,3,4]

[1]Center for Phononics and Thermal Energy Science, School of Physics Science and Engineering, Tongji University, Shanghai 200092, People's Republic of China

[2]Department of Mechanical Engineering, University of Colorado, Boulder, Colorado 80309, USA

[3]Department of Physics, Center for Computational Science and Engineering, and Graphene Research Center, National University of Singapore, Singapore 117546, Republic of Singapore

[4]NUS Graduate School for Integrative Sciences and Engineering, National University of Singapore, Singapore 117456, Republic of Singapore

## Abstract

We theoretically investigate the enhancement of thermoelectric cooling performance in thermoelectric devices made of materials with inhomogeneous thermal conductivity, beyond the usual practice of enhancing thermoelectric figure of merit ZT. The dissipation of Joule heat in such thermoelectric devices is asymmetric which can give rise to better thermoelectric cooling performance. Although the thermoelectric figure of merit and the coefficient-of-performance are only slightly enhanced, both the maximum cooling power and the maximum cooling temperature difference can be enhanced significantly. This finding can be used to increase the heat absorption at the cold end. The asymmetric dissipation of Joule heat also leads to thermal rectification.

* To whom correspondence should be addressed. zhoujunzhou@tongji.edu.cn

There has been great interests in thermoelectric (TE) devices that can directly convert electricity into thermal energy for cooling or heating and can harvest solar and waste heat into electric power [1,2]. The energy conversion efficiency of TE devices is determined by the figure of merit of TE materials [3,4] $ZT = \alpha^2 T/(\rho\lambda)$, where $\alpha$ is the Seebeck coefficient, $T$ is the absolute temperature, $\rho$ is the electrical resistivity, and $\lambda$ is the thermal conductivity which consists of electronic thermal conductivity and lattice thermal conductivity. High ZT materials are desirable for high efficiency TE devices. Even though TE devices have many advantages such as reliability and scalability, the commercial available materials with ZT~1 limits widespread applications of thermoelectrics. Great efforts in enhancing ZT have been made in past decades [5,6,7].

The performance of a TE cooler is evaluated with these three parameters: i). the maximum cooling power $(q_c)_{max}$ that describes the maximum rate at which heat can be absorbed from the cold end, ii) the maximum cooling temperature difference $(\Delta T)_{max}$ which can be reached when the maximum cooling power falls to zero, $(q_c)_{max}=0$; and iii) the maximum coefficient-of-performance (COP) $\phi_{max}$ which is the energy conversion efficiency. There have been many efforts in enhancing the performance of TE coolers through high ZT materials, system engineering [1], and even transient cooling [8,9,10]. In this work, we study the performance of TE devices made of materials with inhomogeneous thermal conductivity.

Assuming *p*- and *n*-type legs have same material properties, we only need to consider a *p*-type branch with length *L* and cross section area *A* as shown in Fig. 1(a) to evaluate device performance [11]. The device is operated with the temperature of $T_1$ and $T_2$ at the cold and hot end, respectively. When an electric current $I$ flows across the device along *x*-direction, heat can be absorbed at rate $q_{ab} = \alpha I T_1$ at the cold end due to Peltier effect. As shown in Fig. 1(a), the absorbed heat can be partially cancelled by the heat leakage due to the temperature difference between the hot and cold ends $q_{DT}$ and the flow of a portion of Joule heat ($q_{Joule} = I^2 R$) generated inside the device where $R$ is the electrical resistance. The net cooling power can then be expressed as

$$q_c = q_{ab} - q_{DT} - \gamma q_{Joule}, \tag{1}$$

where $\gamma$ is defined as inhomogeneity factor of asymmetric Joule heat dissipation. It is rather straightforward that to enhance the device performance $(q_c)_{max}$, $(\Delta T)_{max}$ and $\phi_{max}$ [11,12], one needs to either enhance the Seebeck coefficient $\alpha$ or suppress $q_{DT}$ and $q_{Joule}$.

It is interesting to note that most past studies assume, by default, symmetric flow of the Joule heat to the cold and hot ends, namely, $\gamma = 1/2$ in Eq. (1) [11,12]. However, this assumption is valid only when all the transport coefficients are not spatial-dependent. The factor $\gamma$ can be very different from 1/2 in inhomogeneous materials, which indeed gives rise to a great design freedom to improve the TE cooling performance. Indeed, the devices made of functional graded TE materials (FGTM) with inhomogeneous transport properties was first proposed by Ioffe [13] in 1960 and then be widely studied by many researchers to enhance the device performance [14,15,16,17,18,19]. For example, Bian *et al.* [20] found that an enhancement of $(\Delta T)_{max}$ can be achieved in FGTM with spatial-dependent Seebeck coefficient.

In this Letter, we investigate the performance of TE devices made of inhomogeneous materials with varied transport coefficients. By assuming spatial- and temperature-dependent electrical resistivity $\rho(x,T)$, Seebeck coefficient $\alpha(x,T)$, and thermal conductivity $\lambda(x,T)$, the following equation will be solved to analyze the device performance:

$$\frac{d}{dx}[\lambda(x,T)\frac{dT(x)}{dx}] = -\frac{I^2\rho(x,T)}{A^2} + \frac{I}{A}T(x)\frac{d\alpha(x,T)}{dx}, \quad (2)$$

where $x$ is the distance from the cold end. The boundary conditions are chosen as $T(x=0)=T_1$ and $T(x=L)=T_2$. In Eq. (2), the left term is the divergence of the Fourier heat current, while the first term on the right is the Joule heat generated by an electric current $I$ flowing through the device, and the second term is the Thomson heating or cooling due to the temperature- and spatial-dependent Seebeck coefficient. The temperature profile $T(x)$ can be solved with given $\lambda(x,T)$, $\rho(x,T)$, and $\alpha(x,T)$. The cooling power can be then obtained as $q_c = \alpha(0,T_1)IT_1 - \lambda(0,T_1)A[dT(x)/dx]_{x=0}$ with the temperature profile.

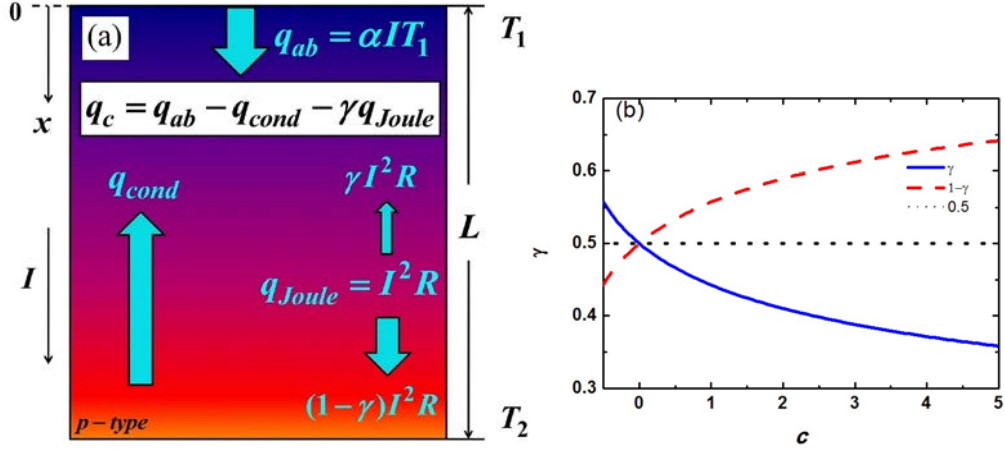

FIG. 1 (color online). (a) Schematic diagram of a TE element with length $L$ and cross-sectional area $A$. The temperature at the cold and hot end are kept at $T_1$ and $T_2$, respectively. The cooling power $q_c$ is the Peltier heat absorbed $q_{ab} = \alpha I T_1$ subtracted by conducted heat due to temperature different between the hot and cold end $q_{DT}$ and a fraction of Joule heat generated inside the TE element $\gamma q_{Joule} = \gamma I^2 R$, where $\alpha$ is the Seebeck coefficient, $R$ is the electrical resistance, and $\gamma$ is the inhomogeneity factor. (b) Fraction of the Joule heat flow to the cold end ($\gamma$) and that flow to the hot end ($1-\gamma$) as a function of parameter $c$ when the inhomogeneous thermal conductivity is $\lambda(x) = \lambda_0 (1 + cx/L)$.

Without losing generality, we study here the enhancement on the cooling performance of TE devices by utilizing the inhomogeneous materials with spatial- and temperature-dependent lattice thermal conductivities. For simplicity, the electrical resistivity and the Seebeck coefficient are set as constant values. Our model can be extended to the case with varied Seebeck coefficients and electrical resistivities. In order to unravel the underlying enhancement mechanism for the cooling performance, the spatial dependence of thermal conductivities and temperature-dependence of thermal conductivities are treated separately. The dependence of $\lambda(T)$ on temperature is a well-known material property. The temperature-dependent $\lambda(T)$ can induce an intrinsic spatial-dependent $\lambda(T(x))$ since the temperature $T(x)$ is spatial-dependent. Materials with explicit spatial-dependent $\lambda(x)$, i.e. through mass gradient and other mechanisms [21,22], has recently been developed to realize thermal rectification effect or thermal diode [23,24,25].

Table I lists several common analytical expressions of spatial- and temperature-dependent thermal conductivities considered in this work. The first example is the inhomogeneous materials with linear spatial-dependent thermal conductivity of $\lambda(x) = \lambda_0(1 + cx/L)$. Here $\lambda_0$ is the reference thermal conductivity at the cold end and the slope $c$ denotes the strength of the spatial dependence or inhomogeneity. The cooling power for this kind of material can then be derived as:

$$q_c = \alpha I T_1 - \beta K_0 \Delta T - \gamma I^2 R, \tag{3}$$

where $\Delta T = T_2 - T_1$, $K_0 = A\lambda_0/L$, and $R = L\rho/A$. Here we have introduced two new parameters of $\beta$ and $\gamma$. The $\beta = c/\ln(1+c)$ is the normalized heat conducted by assuming a homogeneous material with $\lambda_0$, i.e. $q_{DT}/K_0\Delta T$. When $c > 0$, we have $\beta \geq 1$ which means that more heat would be conducted from the hot end to the cold end than $q_{DT} = K_0\Delta T$ due to a much larger effective thermal conductivity. The inhomogeneity factor $\gamma = 1/\ln(1+c) - 1/c$ which denotes the distribution of Joule heat is no longer 1/2. This inhomogeneity factor $\gamma$ can now be tuned by the strength of spatial-dependent thermal conductivity $c$. In the limit of homogeneous case with $c = 0$, the familiar result of $\gamma = 1/2$ can be recovered as expected. Figure 1(b) shows the modulation of $\gamma$ and $1-\gamma$ as a function of parameter $c$. In general, the Joule heat will not flow to the cold and hot ends symmetrically. The Joule heat flow to the cold end $\gamma$ decreases monotonically with increasing $c$. For example, when $c = 13.45$ which means the thermal conductivity varies from $\lambda(0) = \lambda_0$ to $\lambda(L) = 14.45\lambda_0$, $\gamma$ is about 0.3 which is much less than $1 - \gamma = 0.7$. The Joule heat prefers to flow along the direction with increasing thermal conductivities. The discovery of this novel phenomenon enables us to manipulate the Joule heat flow to enhance the cooling performance of TE devices using inhomogeneous thermal conductivities.

The introduction of inhomogeneity modifies the expression of the maximum cooling power $(q_c)_{max}$ which now becomes,

$$(q_c)_{max} = \frac{1}{2\gamma} B - \beta K_0 \Delta T, \tag{4}$$

when the maximum electric current $I_m = \alpha T_1 / 2\gamma R$ is reached, where $B = (\alpha T_1)^2 /(2R)$.

In comparison with the homogeneous thermal conductivity case, when $\lambda(x) = a\lambda_0$ with $a$ as an arbitrary coefficient, there is one more factor $1/2\gamma$ in $I_m$ and in the first term on the right side of Eq. (4), shown in Table I. The maximum cooling temperature difference $(\Delta T)_{max}$ is obtained by setting $(q_c)_{max} = 0$

$$(\Delta T)_{max} = \frac{1}{2\gamma\beta} \frac{B}{K_0}. \tag{5}$$

It is obvious that both $(q_c)_{max}$ and $(\Delta T)_{max}$ can be enhanced when $1/2\gamma > 1$.

In order to calculate the COP written as $\phi = q_c /(\alpha I \Delta T + I^2 R)$ which is the ratio between cooling power and total input power, we now redefine an effective figure of

merit as

$$ZT_M = \frac{1}{\beta} Z_0 T_M, \qquad (6)$$

where $Z_0 = \alpha^2/(K_0 R)$, and $T_M = (1-\gamma)T_1 + \gamma T_2$ is the mean temperature weighted by the inhomogeneity factor $\gamma$. Using such an effective figure of merit and weighted mean temperature and setting $d\phi/dI = 0$, the maximum COP is obtained as

$$\phi_{\max} = \frac{T_1[(1+Z_0 T_M/\beta)^{\frac{1}{2}} + 1 - 2T_M/T_1]}{(T_2 - T_1)[(1+Z_0 T_M/\beta)^{\frac{1}{2}} + 1]}. \qquad (7)$$

In the limit of homogeneous case when $c=0$, i. e. $\lambda(x) = \lambda_0$, the familiar results of $\beta = 1$, $T_M = (T_1+T_2)/2$, and the conventional expression of $\phi_{\max}$ with homogeneous thermal conductivity are recovered [12].

We perform the numerical calculations based on the above mentioned model for a TE element with $L$=5 mm and $A$=4mm$^2$. The typical material properties of $p$-type BiSbTe alloy [26] have been adopted as follows: the Seebeck coefficient $\alpha = 220\,\mu\text{V/K}$, the electrical resistivity $\rho = 10^{-5}\,\Omega\text{m}$, and $\lambda_0 = 1.7\,\text{W/(m}\cdot\text{K)}$. The temperature at the hot end is fixed to be $T_2 = 300\,\text{K}$ for all calculations. The temperature at the cold end is chosen to be $T_1 = 290\,\text{K}$ for the calculations of $(q_c)_{\max}$ and $\phi_{\max}$. In the calculation of $(\Delta T)_{\max}$, $T_1$ is obtained by solving Eq. (5) self-consistently.

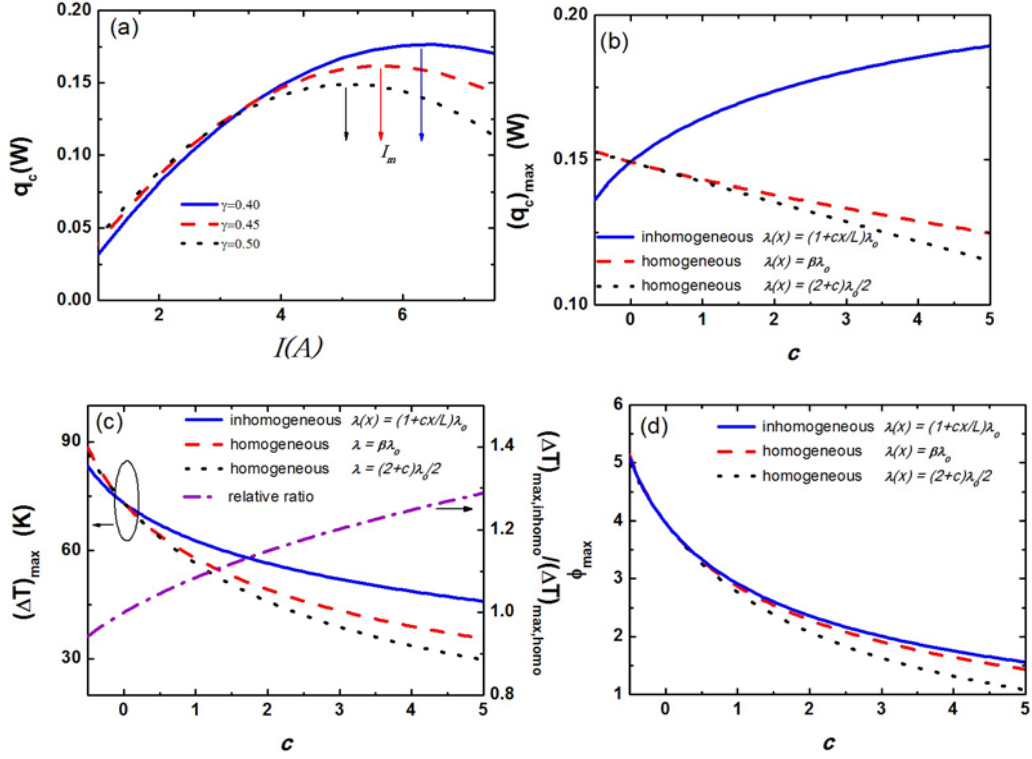

FIG. 2 (color online) Enhancement of $(q_c)_{max}$, $(\Delta T)_{max}$, and $\phi_{max}$ by inhomogeneous materials with linear spatial-dependent thermal conductivities $\lambda(x) = \lambda_0(1+cx/L)$. (a) $q_c$ as a function of electric current $I$ for different $\gamma$. (b), (c) and (d), $(q_c)_{max}$, $(\Delta T)_{max}$, and $\phi_{max}$ as a function of parameter $c$ (blue solid curve), respectively. For comparison, $(q_c)_{max}$, $(\Delta T)_{max}$, and $\phi_{max}$ with homogeneous thermal conductivities, $\lambda = \beta\lambda_0$ (red dashed curve) and $\lambda = \lambda_0(2+c)/2$ (black dotted curve), are also plotted. Relative ratio between $(\Delta T)_{max}$ with inhomogeneous thermal conductivity and $(\Delta T)_{max}$ with homogeneous thermal conductivity $\lambda = \beta\lambda_0$ is plotted in (c).

Figure 2(a) shows the cooling power $q_c$ as a function of the electric current $I$ with linear spatial-dependent thermal conductivities $\lambda(x) = \lambda_0(1+cx/L)$ when $c=0$, 0.85, and 2.4. The corresponding inhomogeneity factors are $\gamma = 0.5$ for $c=0$, $\gamma = 0.45$ for $c=0.85$, and $\gamma = 0.4$ for $c=2$, respectively. The maximum electric current $I_m$ shifts from 5.1A to 6.4A when $\gamma$ decreases from 0.5 to 0.4 since $I_m \propto 1/\gamma$ as shown in Table I. In the mean time, the maximum cooling power $(q_c)_{max}$ increases from 0.149W to 0.177W. When $c$ increases, both the maximum electric current $I_m$ and normalized conducted heat $\beta$ are increased. When the increase of $\alpha I_m T_1/2$ overcomes the increase of $\beta K_0 \Delta T$, the overall effect is the enhancement of $(q_c)_{max}$, by recalling Eq. (4).

Figures 2(b)-2(d) plot the maximum cooling power $(q_c)_{max}$, the maximum cooling temperature difference $(\Delta T)_{max}$, and the maximum COP $\phi_{max}$ as a function of $c$ with inhomogeneous thermal conductivity $\lambda(x) = \lambda_0(1+cx/L)$, respectively. For comparison, $(q_c)_{max}$, $(\Delta T)_{max}$, and $\phi_{max}$ with homogeneous thermal conductivities, $\lambda = \beta\lambda_0$ and $\lambda = \lambda_0(2+c)/2$, are also presented. These two cases of homogeneous thermal conductivities are chosen for comparison because: i) both thermal conductivities $\lambda = \beta\lambda_0$ and $\lambda(x) = \lambda_0(1+cx/L)$ result in the same normalized conducted heat $\beta$; ii) $\lambda = \lambda_0(2+c)/2$ is the mean value of linear spatial-dependent thermal conductivities over the length of device, i. e. $\lambda_0(2+c)/2 = \int_0^L \lambda_0(1+cx/L)dx/L$.

Figure 2(b) shows that the maximum cooling power with inhomogeneous thermal conductivity in Eq. (4) increases as $c$ increases. When $c > 0$, we find that $(q_c)_{max}$ is significantly larger than the maximum cooling power with homogeneous thermal conductivities which are noted as $(q_c)'_{max} = B - \beta K_0 \Delta T$ for $\lambda(x) = \beta\lambda_0$ and $(q_c)''_{max} = B - K_0 \Delta T(2+c)/2$ for $\lambda(x) = \lambda_0(2+c)/2$. These three maximum cooling powers satisfy the inequality $(q_c)_{max} > (q_c)'_{max} > (q_c)''_{max}$ since $1/2\gamma > 1$ and $\beta < (2+c)/2$. For example, $(q_c)_{max}$ is 0.189W which is larger than $(q_c)'_{max} = 0.125$W and $(q_c)''_{max} = 0.115$W when $c = 5$. On the contrary, negative $c$ results in a smaller $(q_c)_{max}$ with inhomogeneous thermal conductivities than both $(q_c)'_{max}$ and $(q_c)''_{max}$ with homogeneous thermal conductivities since $1/2\gamma < 1$ when $c < 0$.

Figure 2(c) shows that the maximum cooling temperature difference $(\Delta T)_{max}$ with inhomogeneous thermal conductivity in Eq. (5) decreases with increasing $c$. The reason is that the factor $2\beta\gamma$ increases with increasing $c$. When $c > 0$, we find that $(\Delta T)_{max}$ is larger than that with homogeneous thermal conductivities which are noted as $(\Delta T)'_{max} = B/(\beta K_0)$ for $\lambda = \beta \lambda_0$ and $(\Delta T)''_{max} = B/[K_0(2+c)/2]$ for $\lambda = \lambda_0(2+c)/2$. These three maximum cooling temperature differences satisfy the inequality $(\Delta T)_{max} > (\Delta T)'_{max} > (\Delta T)''_{max}$ since $1/2\gamma > 1$ and $1/\beta > 2/(2+c)$. For example, $(\Delta T)_{max}$ is 46K which is larger than $(\Delta T)'_{max} = 36$K and $(\Delta T)''_{max} = 30$K when $c = 5$. On the contrary, negative $c$ results in a smaller $(\Delta T)_{max}$ with inhomogeneous thermal conductivity than both $(\Delta T)'_{max}$ and $(\Delta T)''_{max}$ with homogeneous thermal conductivity since $1/2\gamma < 1$ when $c < 0$.

Figure 2(d) shows the maximum COP with inhomogeneous thermal conductivity decreases with increasing $c$. The reason is that $\phi_{max}$ is proportional to the figure of merit, $\phi_{max} \sim ZT_M = Z_0 T_M / \beta$ as shown in Eq. (6), which decreases as $c$ increases. When $c > 0$, we find that $\phi_{max}$ is larger than that with homogeneous thermal conductivities which are noted as $\phi'_{max}$ and $\phi''_{max}$ for $\lambda = \beta \lambda_0$ and $\lambda = \lambda_0(2+c)/2$, respectively. The effective figure of merit and mean temperature are $Z_0/\beta$ and $(T_1+T_2)/2$ in the calculation of $\phi'_{max}$, $2Z_0/(2+c)$ and $(T_1+T_2)/2$ in the calculation $\phi''_{max}$. The weighted mean temperature $T_M$ in the calculation of $\phi_{max}$ is slightly smaller than $(T_1+T_2)/2$. For instance, when $c=5$ which leads to $\gamma = 0.36$, $T_M = 293.6K$ is $1.4K$ smaller than $(T_1+T_2)/2 = 295K$. The relative difference between them is below 0.5%. Such tiny difference makes $\phi_{max}$ slightly larger than $\phi'_{max}$. Moreover, $\phi_{max}$ is larger than $\phi''_{max}$ because that $Z_0/\beta > 2Z_0/(2+c)$ results in a lager $Z$. On the contrary, negative $c$ results in a smaller $\phi_{max}$ with inhomogeneous thermal conductivities than $\phi'_{max}$ and $\phi''_{max}$ with homogeneous thermal conductivities since $Z_0/\beta < 2Z_0/(2+c)$ and $T_M > (T_1+T_2)/2$ when $c<0$.

Besides homogeneous thermal conductivity $\lambda = a\lambda_0$ and linear spatial-dependent thermal conductivity $\lambda(x) = \lambda_0(1+cx/L)$ cases, we also investigate TE cooling performance of the explicit spatial-dependent thermal conductivity with power law dependence $\lambda(x) = \lambda_0(x/L)^d$ $(d<1)$ and exponential dependence $\lambda(x) = \lambda_0 \exp[gx/L]$ as shown in Table I. One can see that $I_m$, $(\Delta T)_{max}$, and $Z$ of power law and exponential spatial-dependent thermal conductivities have the same expressions as that of linear spatial-dependent thermal conductivities except that the expressions of $\beta$ and $\gamma$ are changed. More detailed numerical results are given in the Supplemental Material [27].

TABLE I. Expressions of the normalized conducted heat $\beta$, the inhomogeneity

factor $\gamma$, the maximum electric current $I_m$, the maximum cooling temperature difference $(\Delta T)_{max}$, and the effective figure of merit $ZT_M$ for different explicit and intrinsic spatial-dependent thermal conductivities $\lambda(x,T)$. Where $\beta = q_{DT}/(K_0 \Delta T)$, $B = (\alpha T_1)^2/(2R)$, $Z_0 = \alpha^2/(K_0 R)$, $T_M = (1-\gamma)T_1 + \gamma T_2$. Remember that $(q_c)_{max} = \alpha T_1 I_m/2 - q_{DT}$ and $\phi_{max}$ increases with increasing $Z$.

| $\dfrac{\lambda(x,T)}{\lambda_0}$ | $\beta$ | $\gamma$ | $\dfrac{I_m}{\alpha T_1/R}$ | $\dfrac{(\Delta T)_{max}}{B/K_0}$ | $\dfrac{Z}{Z_0}$ |
|---|---|---|---|---|---|
| $a$ | $a$ | $1/2$ | $1$ | $\dfrac{1}{\beta}$ | $\dfrac{1}{\beta}$ |
| $1 + cx/L$ | $\dfrac{c}{\ln(1+c)}$ | $\dfrac{1}{\ln(1+c)} - \dfrac{1}{c}$ | $\dfrac{1}{2\gamma}$ | $\dfrac{1}{2\gamma\beta}$ | $\dfrac{1}{\beta}$ |
| $(x/L)^d$ | $1 - d$ | $\dfrac{1-d}{2-d}$ | $\dfrac{1}{2\gamma}$ | $\dfrac{1}{2\gamma\beta}$ | $\dfrac{1}{\beta}$ |
| $\exp(gx/L)$ | $\dfrac{g}{(1-e^{-g})}$ | $\dfrac{1}{g} - \dfrac{1}{e^g - 1}$ | $\dfrac{1}{2\gamma}$ | $\dfrac{1}{2\gamma\beta}$ | $\dfrac{1}{\beta}$ |
| $(T/T_0)^{-1}$ | $\dfrac{T_0(\ln T_2 - \ln T_1)}{\Delta T}$ | $\dfrac{1}{2}$ | $1$ | N/A | N/A |
| $(T/T_0)^h$ | $\dfrac{T_2^{h+1} - T_1^{h+1}}{(h+1)T_0^h \Delta T}$ | $\dfrac{1}{2}$ | $1$ | N/A | N/A |
| $1 + bT/T_0$ | $1 + b\dfrac{T_1 + T_2}{2T_0}$ | $\dfrac{1}{2}$ | $1$ | $\dfrac{1}{\beta}$ | $\dfrac{1}{\beta}$ |

From Fig. 2 and Table I, the results with explicit spatial-dependent thermal conductivity can be briefly described as follows: i) the maximum cooling power and the maximum cooling temperature difference can be greatly enhanced while the maximum COP is only slightly enhanced in TE device; ii) to enhance the cooling performance, the thermal conductivity close to the cold end should be smaller than the thermal conductivity close to hot end, which results in a smaller fraction of the Joule heat flow towards the cold end, as noted by $\gamma < 1/2$ when $c > 0$, $d > 0$, and $g > 0$ as shown in Table I.

Table I also summarizes the results with temperature-dependent, or intrinsic spatial-dependent, thermal conductivities with power law temperature-dependence $\lambda(T) = \lambda_0 (T/T_0)^h$ ($h = -1$ and $h \neq -1$ are present separately in Table I) and linear temperature-dependence $\lambda(T) = \lambda_0 (1 + bT/T_0)$ where $T_0$ is the room temperature. The detailed numerical results can be found in Supplemental Material [27]. One important observation is that the intrinsic spatial-dependent thermal conductivities due to its dependence on temperature do not lead to the asymmetric dissipation of Joule heat. In other words, $\gamma$ is always equal to 1/2. The Joule heat flowing towards the cold end is exactly the same as the case with homogeneous thermal conductivity. Therefore the maximum electric current $I_m$ is the same as that with homogeneous thermal conductivity. Only the normalized conducted heat $\beta$ is modified. Furthermore, there is no simple explicit forms of $(\Delta T)_{\max}$ and $Z$ for the case with $\lambda(T) = \lambda_0 (T/T_0)^h$ which are noted as N/A in Table I.

We believe that there is a fundamental difference between the explicit spatial-dependent thermal conductivities case and the temperature-dependent thermal conductivities case. The physical explanation is that space inversion symmetry is broken for explicit spatial-dependent thermal conductivities, but conserved for temperature-dependent thermal conductivities. If we swap the boundary condition, $T_1 \leftrightarrow T_2$, the heat transport process and temperature profile after the reversion is exactly the same as that before the reversion. This might also be the reason why there is no thermal rectification effect for homogeneous materials with temperature-dependent thermal conductivities. Our earlier research shows that it is crucial to utilize some kind of symmetry breaking mechanism to realize a thermal diode [24,25].

Since the inversion symmetry is broken by the spatial-dependent thermal conductivities, the resulted asymmetric Joule heat flow can also be used for novel design of thermal diodes. In particular, without considering the Peltier effect, i.e. $\alpha \to 0$, the heat current flowing out of the device changes from $q^+ = \beta K_0 \Delta T + \gamma I^2 R$ to $q^- = \beta K_0 \Delta T + (1-\gamma) I^2 R$ if the boundary condition is swapped ($T_1 \leftrightarrow T_2$). Therefore, the thermal rectification factor can be derived as [25]:

$$R_f = \frac{q^+ - q^-}{q^-} = \frac{(2\gamma - 1)}{\beta/\eta + (1-\gamma)}, \tag{8}$$

where $\eta = I^2 R / (K_0 \Delta T)$ denotes the normalized Joule heat. The rectification factor $R_f$ varies from $-1 \sim \infty$ for the ideal thermal diode [24,28].

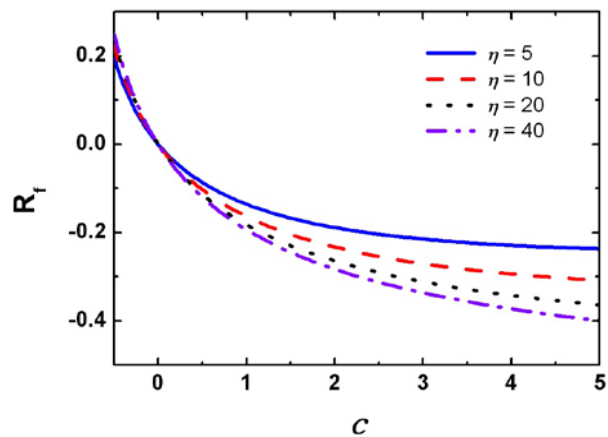

FIG. 3 (color online) Thermal rectification factor $R_f$ versus parameter $c$ with inhomogeneous thermal conductivity $\lambda(x) = \lambda_0(1+cx/L)$ for different normalized Joule heat $\eta$.

It is obvious that any deviation from 1/2 for the inhomogeneity factor $\gamma$ will induce a finite thermal rectification effect for nonzero Joule heat. Figure 3 shows the thermal rectification factor $R_f$ as a function of parameter $c$ of the linear spatial-dependent thermal conductivities with $\lambda(x) = \lambda_0(1+cx/L)$ for different normalized Joule heat $\eta$. We find that the $R_f$ is positive when $c<0$ and negative when $c>0$. Larger $|c|$ leads to an enhancement of $|R_f|$ that means stronger rectification. $|R_f|$ increases with increasing normalized Joule heat $\eta$ since the contribution of Joule heat to total heat current is enlarged.

To summarize, we have discovered that thermoelectric cooling performance can be significantly enhanced through the manipulation of Joule heat flow with explicit spatial-dependent inhomogeneous thermal conductivity. The flow of Joule heat towards the cold end can be suppressed when the thermal conductivity near the cold end is smaller than that near the hot end. We found that the maximum cooling power and the maximum cooling temperature difference can be significantly enhanced while the coefficient-of-performance is slightly enhanced. The intrinsic spatial-dependent thermal conductivity due to its temperature dependence cannot lead to such enhancement. Our findings suggest that the materials with inhomogeneous thermal conductivity used for thermal rectifier/diode can be also used to improve the performance of thermoelectric cooling, which in turn enriches the applications of thermal rectifier [25].

It should be pointed out that materials with inhomogeneous thermal conductivity can be now be readily achieved with nanotechnology. For example, the inhomogeneous nanotube [22], thin diamond film in which the inhomogeneity is due to spatially varying disorder associated with nucleation and grain coalescence [29], and thermal rectifier with pyramid shaped $LaCoO_3/La_{0.7}Sr_{0.3}CoO_3$ [30]. We expect that our investigation will inspire many follow-up works in realizing inhomogeneous thermal conductivity and wide-spread applications of thermal rectifiers.

**Acknowledgments**

TL, JZ, NL, and BL are supported by the NSF China, with grant No. 11334007. NL is also supported by Shanghai Rising-Star Program with grant No. 13QA1403600. JZ is also supported by the program for New Century Excellent Talents in Universities grant no. NCET-13-0431. RY acknowledges the support from NSF and AFOSR from the United States.

performance with power law and exponential spatial-dependent thermal conductivities; linear and power law temperature-dependent thermal conductivities.